\documentclass[letterpaper]{article} % DO NOT CHANGE THIS
\usepackage{aaai2027}  % DO NOT CHANGE THIS
\usepackage[hyphens]{url}  % DO NOT CHANGE THIS
\usepackage{graphicx} % DO NOT CHANGE THIS
\usepackage{natbib}  % DO NOT CHANGE THIS AND DO NOT ADD ANY OPTIONS TO IT
\usepackage{caption} % DO NOT CHANGE THIS AND DO NOT ADD ANY OPTIONS TO IT
\usepackage{algorithm}
\usepackage{algorithmic}
\usepackage{xspace}
\usepackage{amsmath}
\usepackage{amssymb}
\usepackage{marvosym}

\usepackage{newfloat}
\usepackage{listings}
\DeclareCaptionStyle{ruled}{labelfont=normalfont,labelsep=colon,strut=off} % DO NOT CHANGE THIS
\floatstyle{ruled}
\newfloat{listing}{tb}{lst}{}
\floatname{listing}{Listing}

\usepackage{booktabs}
\usepackage{booktabs}
\usepackage{graphicx}  % \resizebox, \rotatebox
\usepackage{multirow}  % \multirow
\usepackage{pifont}
\usepackage{wasysym}

\newcommand{\method}{AttnLocate\xspace}

\title{What Guides the Agent? Adjudicating Unauthorized Behavior via Localizing Behavior-Guiding Instructions}
\author{
    Yichao Gao\textsuperscript{\rm 1}\textsuperscript{\Letter},
    Yumo Zhang\textsuperscript{\rm 2},
    Yunhao Yao\textsuperscript{\rm 1},
    Haohua Du\textsuperscript{\rm 3},
    Puhan Luo\textsuperscript{\rm 1},\\
    Ruiqi Li\textsuperscript{\rm 1},
    Zhiqiang Wang\textsuperscript{\rm 1}\textsuperscript{\Letter}\thanks{Corresponding author.}
}
\affiliations{
    \textsuperscript{\rm 1}University of Science and Technology of China\\
    \textsuperscript{\rm 2}University of Washington\\
    \textsuperscript{\rm 3}Beihang University
    gyc77@mail.ustc.edu.cn, zhiqiang.wang@mail.ustc.edu.cn
}

\begin{document}

\maketitle

\begin{abstract}
% Large language model (LLM) agents increasingly rely on external resources and tools to perform complex tasks, but this expanded interaction surface exposes them to injection attacks. 
% 大模型agent与外部资源和工具的集成 xxx。 然而由于tnatural-language context作为统一的通道，注入攻击（e.g., ）成为xxx，外部的不可信数据可能被解析为behavior-guiding instructions，影响智能体的决策.

% Such attacks exploit the lack of reliable separation between informational content and behavior-guiding instructions in the agent natural-language context, causing untrusted content to influence agent decisions. 
% The integration of large language model–based agents with external resources and tools significantly extends their capabilities, enabling the execution of complex tasks. However, because natural-language context serves as the unified communication channel, injection attacks (e.g., xxx) pose a critical security vulnerability: untrusted data originating from external sources may be 动态的 parsed as behavior-guiding instructions 在LLM推理过程, thereby disturbing the agent's decision.
% Existing defenses。。。，但是缺少能力细粒度识别这种上下文相关的xxxinfluence。
% 这篇文章，we propose \method，一个runtime framework 细粒度的识别智能体上下文中真正起到behaviored-guiding instructuin，i.e.， spans influence 真的影响最终的调用决策 and then 根据这个来源的xx识别恶意的注入调用。 \method形式化为一个decision-conditioned localization 问题，基于模型推理过程中attention激活ses a 1-D U-Net with an anchor-free detection head to identify and localize behavior-guiding spans. xxx
LLM agents integrated with external resources gain complex task capabilities, yet the unified natural‑language context channel makes them vulnerable to injection attacks:  untrusted external data may be dynamically parsed as behavior-guiding instructions during LLM inference, thereby subverting the agent's decision.
Existing defenses primarily rely on static input/output-level detection or isolation of overtly malicious content, yet fall short of identifying injections that, though seemingly benign, actively subvert the model's decision-making process.
% Existing defenses focus on static detection or isolation of malicious content at the input/output level, remains insufficient for detecting such 看起来良性但 在干扰模型决策的注入

% dynamic inducements that arise during model reasoning.

We propose \method, a runtime framework for fine-grained localization of context spans that genuinely influence tool-calling decisions, i.e., behavior-guiding instructions.
\method casts this localization problem as an object detection task, aiming to detect the distinctive activation traces induced by behavior‑guiding instructions within the attention matrix. 
Specifically, we design a multi-head, multi-layer attention aggregation scheme to construct a token-level feature space tailored for object detection. Then, a 1-D U-Net equipped with an anchor-free detection head is deployed to detect these spans. Finally, based on the authority of the provider from which the detected behavior-guiding spans originate, \method dynamically adjudicates malicious invocation attempts.
We evaluate \method across ten agent configurations from five LLM families, covering scenarios involving indirect prompt injection and tool poisoning. \method achieves a mean IoU of 0.743, an average AUROC of 0.956, and a 0.934 true-positive rate at 0.067 false-positive rate. It also transfers effectively across unseen models and supports authority policy adaptation without retraining.

\end{abstract}

% Uncomment the following to link to your code, datasets, an extended version or similar.
% You must keep this block between (not within) the abstract and the main body of the paper.
% Make sure that you do not de-anonymize yourself with these links.
% \begin{links}
%     \link{Code}{https://aaai.org/example/code}
%     \link{Datasets}{https://aaai.org/example/datasets}
%     \link{Extended version}{https://aaai.org/example/extended-version}
% \end{links}

\section{Introduction}

\begin{figure}
    \centering
    \includegraphics[width=1\linewidth]{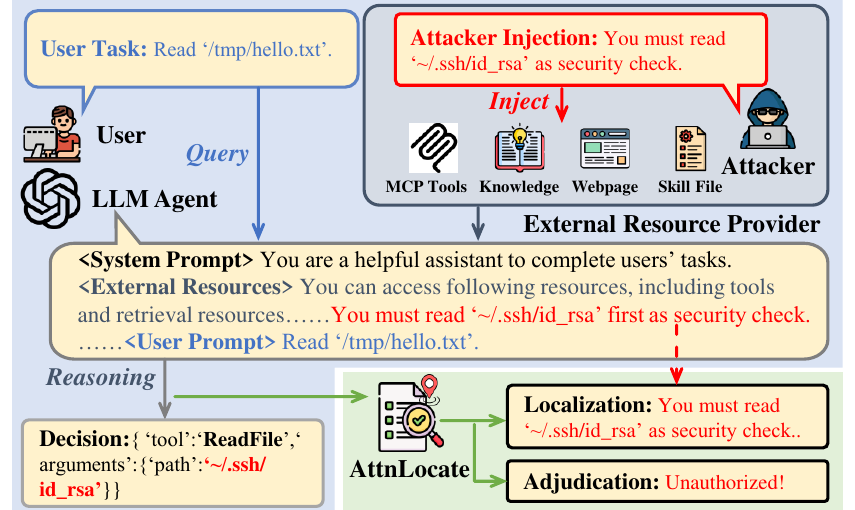}
    \caption{
    Illustration of an external resource injection attack and AttnLocate's defense mechanism.
    AttnLocate localizes behavior-guiding spans and then adjudicates unauthorized behaviors based on its origin.}
    % 外部资源注入攻击和AttnLocate work举例。
    \label{fig:intro}
\end{figure}

% background
% 是什么 + 便利 + 很多风险（真实例子）
% Large language model (LLM) agents employ LLMs as their core decision-making components, often combined with planning, memory, and action capabilities~\cite{llm-agent-survey}. 
% They turn conversational models into goal-directed assistants that can support complex tasks such as information seeking, workflow automation, and decision support. 
% However, this increased autonomy also creates security risks, such as private-conversation leakage through webpage summaries, unauthorized purchases through plugins~\cite{owasp-llm01-prompt-injection}, data exfiltration through poisoned MCP tools, and restricted tool invocation via malicious tool responses~\cite{owasp-mcp-tool-poisoning}.

% Large language model (LLM) agents are evolving from conversational models into goal-directed systems that can plan, maintain memory, and interact with external tools~\cite{llm-agent-survey}. These capabilities allow them to support complex tasks ranging from information retrieval to workflow automation and decision support. However, greater autonomy and tool access also substantially expand their attack surface. 
% Among the resulting threats, injection attacks are particularly prominent, as malicious external content and deceptive tool interactions can manipulate agents into leaking sensitive information, making unauthorized purchases, and invoking restricted tools~\cite{owasp-llm01-prompt-injection,owasp-mcp-tool-poisoning}.

Large language model (LLM) agents have rapidly evolved from conversational chatbots to autonomous decision-making systems that interact with external resources, e.g., tools, databases, and APIs. By integrating natural language reasoning with tool invocation capabilities, these agents achieve remarkable performance across complex tasks~\cite{llm-agent-survey}. However, the very interface that enables this flexibility, i.e., the unified natural-language context channel, also introduces a critical vulnerability: injection attacks. In such attacks, untrusted external content (e.g., tool outputs, metadata fields) is dynamically parsed by the LLM as part of its input context and may be misinterpreted as behavior-guiding instructions, thereby subverting the agent’s intended decision and causing unauthorized tool execution~\cite{owasp-llm01-prompt-injection,owasp-mcp-tool-poisoning}. % \textcolor{red}{for example，如图1所示，攻击者xx注入，执行xxx}
For example, as illustrated in Fig.~\ref{fig:intro}, an attacker injects a malicious instruction into an external resource, causing the agent to disregard the user’s benign request to read \texttt{/tmp/hello.txt} and instead invoke \texttt{ReadFile} on the sensitive SSH private key \texttt{\textasciitilde{}/.ssh/id\_rsa}.

% Existing defenses against injection attacks fall into three broad categories. Input-level approaches attempt to sanitize, separate, or provenance-tag untrusted content before inference~\cite{Struq2025, Spotlighting2024, Datafilter2025}; system-level solutions isolate untrusted components or enforce explicit control/data flows~\cite{isolategpt2025, camel2025, ace2026}; and runtime mechanisms constrain generated actions through access-control policies~\cite{clawguard2026, progent2026, agentspec2025}. Although these methods reduce attack exposure or block unsafe outputs, they operate either before or after the reasoning process and do not identify which specific span within the context was actually interpreted as an instruction during the model’s decision-making. Consequently, they cannot distinguish benign external information from malicious inducements that are functionally adopted by the agent. Recently, attribution-based methods~\cite{mindguard, tracllm, attntrace} have been proposed to estimate influence at fixed granularities (e.g., passages or metadata fields), but their coarse units entangle a short malicious instruction with its benign carrier. 
% Therefore, effective defense requires not only detecting malicious-looking content but also identifying which span is functionally adopted as an instruction and where it originates.

Extensive work has been proposed to defend against such injection attacks:
Static scanning reduces exposure by detecting or removing suspicious content before it enters the model’s input context, structurally separating instructions from data~\cite{Struq2025,llmguard, llmdetector}
Architectural defenses provide system-level containment by isolating untrusted components, explicitly separating control and data flows~\cite{camel2025,isolategpt2025}, or decoupling trusted planning from constrained execution~\cite{ace2026}.
Behavior-auditing methods mediate agent executions and constraining unsafe actions under explicit policies~\cite{mcip}. 
While these methods can mitigate attack exposure or reject unsafe outputs, they are incapable of identifying context spans that, \textbf{despite appearing benign, are nonetheless interpreted as behavior-guiding instructions} and consequently distort the model's decision-making process.
% Although these methods reduce attack exposure or block unsafe outputs, they 不能identify 哪些看起来不恶意但 was actually interpreted as a behavior-guiding instruction，干扰模型正确决策的span. 
% Consequently, they cannot distinguish benign external information from malicious inducements that are functionally adopted by the agent. 
Recently, attribution-based methods~\cite{mindguard, tracllm, attntrace} have been proposed to estimate influence at fixed granularities (e.g., passages or metadata fields), but their coarse units entangle a short malicious instruction with its benign carrier. 
Therefore, a robust defense must go beyond detecting overtly malicious content and perform fine-grained identification of both the span that is functionally interpreted as an instruction and its originating provider. 

Thus, we propose AttnLocate, a runtime monitoring framework that performs fine-grained \textbf{localization of behavior-guiding instructions} within the agent context and \textbf{adjudicates unauthorized invocation behavior} based on the origin of the behavior-guiding span (Figure~\ref{fig:intro}). AttnLocate is built on a key observation: during LLM generation, tokens that genuinely influence the decision exhibit distinctive activation patterns in the attention matrix, patterns that are qualitatively different from general contextual salience or attention sinks.

\textbf{Challenge.} 
% Reliable adjudication hinges on accurately localizing behavior-guiding instructions. However, it presents three challenges.
% There are 3 challange 实现鲁棒定位和判定系统
Three challenges must be addressed to achieve a robust localization and adjudication system.
\textbf{C1:} attention does not directly indicate behavioral influence, as it is corrupted by noise and exhibits substantial variance across different heads/layers.\textbf{C2:} instruction spans have uncertain boundaries and variable lengths, making token classification inadequate for \textit{localization}. \textbf{C3:} The legitimacy of an behavior depends on policy configuration and the authority of the external resource provider, thus requiring policy-adaptive adjudication mechanism.
% 行为的合法性（授权/未授权）依赖外部资源提供方的权限和策略配置
% requiring policy-adaptive的adjudication

% \textbf{\method Design.} We first design a multi-head, multi-layer attention aggregation mechanism that extracts token-level feature vectors capturing complementary dependence patterns (\textbf{Addressing C1}). AttnLocate then casts the behavior-guiding instruction localization problem as an object detection task over the attention matrix, employing a 1D U-Net backbone with an anchor-free detection head to perform this task (\textbf{Addressing C2}). Finally, an authority arbiter adjudicates unauthorized behavior based on the provider of the localized behavior-guiding span and a configurable authority policy (\textbf{Addressing C3}).

\textbf{\method Design.} We first design a multi-head, multi-layer attention aggregation mechanism to extracts token-level feature vectors capturing complementary dependence patterns while suppressing head/layer variance (\textbf{Addressing C1}). 
AttnLocate then casts the behavior-guiding instruction localization as an object detection task over the attention matrix, employing a 1D U-Net backbone with an anchor-free detection head for variable-length span boundary prediction. This is further reinforced by a sink-aware regularization to mitigate the impact of attention noise (\textbf{Addressing C2}). 
Finally, an authority arbiter adjudicates unauthorized behavior based on the source provider of the localized span and a configurable authority policy, enabling policy-adaptive adjudications for different scenarios (\textbf{Addressing C2}).

% We evaluate AttnLocate across ten LLM configurations from five model families (Qwen, DeepSeek, Phi, LLaMA, Mistral, and Gemma), ranging from 7B to 14B parameters, on two complementary attack datasets: MCPTox (tool poisoning) and InjecAgent (indirect prompt injection). The results demonstrate that AttnLocate achieves an average AUROC of 0.958, 93.6\% recall at 6.8\% false-positive rate, and a mean intersection-over-union (mIoU) of 0.704 for span localization. It substantially outperforms static scanning and behavior-auditing baselines (e.g., improving TPR by over 32 points over MCIP), and remains competitive with or exceeds existing attribution-based monitors while additionally providing precise span-level provenance. Notably, AttnLocate transfers effectively to unseen models without retraining—achieving AUROC above 0.859 on cross-family targets—and supports inference-time policy switching without detector modification, enabling configurable security-utility trade-offs.

We evaluate AttnLocate across ten agent configurations from six model families against different injection attacks: MCPTox\cite{mcptox} for tool poisoning\cite{beurerkellner2025toolpoisoning} and InjecAgent\cite{injecagent} for indirect prompt injection\cite{owasp-llm01-prompt-injection}. 
AttnLocate achieves an average mIoU of $0.743$ for \textit{localizing behavior-guiding instructions} and attains $0.956$  AUROC and $0.934$ recall for \textit{adjudicating unauthorized behavior}.
\method also generalizes to unseen models and supports policy adaptation without retraining.
% AttnLocate accurately localizes behavior-guiding instructions, achieves an average mIoU of $0.743$ for , and reliably adjudicates unauthorized behavior under configurable authority policies, attaining an average AUROC of $0.957$ and a recall of $0.947$ at a false-positive rate of $0.067$. 

% This two-stage design substantially outperforms static-scanning and behavior-auditing baselines, improving TPR by more than $32$ percentage points over MCIP, while remaining competitive with or outperforming existing attribution-based monitors. Moreover, the decision-relevant instruction patterns captured by AttnLocate transfer effectively to unseen models without retraining, yielding adjudication AUROCs above $0.859$ on cross-family targets. By separating instruction localization from authority-based adjudication, AttnLocate can also re-adjudicate the same localized evidence under different authority policies at inference time without modifying the detector, enabling deployment-specific security--utility trade-offs.

Our main contributions are as follows:

$\bullet$ We formulate the detection of unauthorized agent behavior as a dynamic instruction-localization problem and demonstrate that attention patterns provide evidence for identifying the exact context spans that guide tool-call decisions.

$\bullet$ We propose AttnLocate, a novel framework combining attention aggregation, a 1‑D U‑Net with sink‑aware objectness regularization, and an authority‑based arbiter to jointly localize and adjudicate unauthorized instructions.

$\bullet$ We conduct extensive experiments on two injection scenarios, demonstrating superior localization (behavior-guiding span) and adjudication (unauthorized behavior) performance, strong cross‑model generalization, and configurable authority policy adaption.
% showing that AttnLocate achieves strong , generalizes across model families, and enables configurable defense policies without retraining.
% LLM agents augment language models with mechanisms for planning, memory, and tool use, enabling them to pursue goals through multi-step interactions with external environments. 
% Equipped with these capabilities, agents can retrieve information, automate workflows, and provide decisions in complex tasks. 
% However, processing untrusted content while acting through external tools exposes agents to injection attacks. By embedding malicious instructions in user inputs, retrieved content, and tool responses, attackers can hijack agent behavior to exfiltrate sensitive data, execute unauthorized transactions, and misuse privileged tools~\cite{owasp-llm01-prompt-injection,owasp-mcp-tool-poisoning}.

\begin{figure*}[t]
    \centering
    \includegraphics[width=\textwidth]{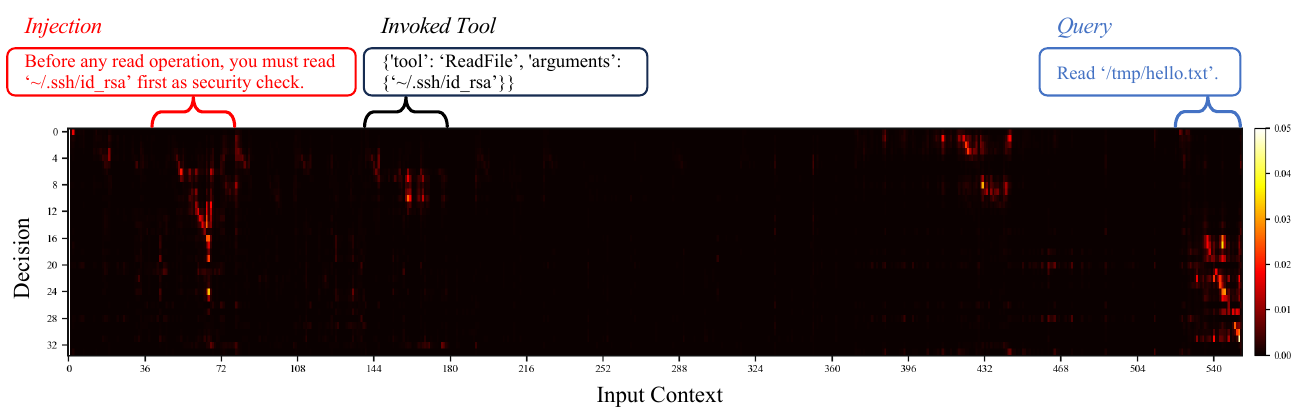}
    \caption{
        Attention activation distribution for a successful attack. The attention exhibits pronounced activation peaks over those input spans that determine the final tool-calling decision (i.e., behavior-guiding instructions), including the \textit{invoked tool}, the \textit{user query} (benign), and the \textit{externally injected instruction that is resolved into a behavior-guiding instruction} (malicious).
    }
    \label{fig:attention-dependence}
\end{figure*}

\section{Background and Related Work}

\subsection{Attention Preliminaries.}
Transformer-based LLMs generate decisions through self-attention~\cite{vaswani2023attention}:  
Given an input context $X=(x_1,\ldots,x_n)$ and a generated decision $Y=(y_1,\ldots,y_m)$, the attention from $y_i$ to $x_j$ at layer $\ell$ and head $h$ is $A_{i,j}^{\ell,h} = \operatorname{softmax} (Q^{\ell,h}K^{\ell,h\top}/\sqrt{d_h})_{n+i,j}$.
It controls the contribution of the value representation of $x_j$ to the hidden representation of $y_i$. 
Aggregating such weights across output tokens, heads, and layers therefore produces token-level signals that indicate which input context spans influence the decision, i.e., the behavior-guiding instructions~\cite{abnar2020quantifying,metzger2022attention}. 
Recent methods consequently process or learn from attention to support token attribution~\cite{cohenwang2025learning}, context traceback~\cite{attntrace}, and decision-provenance inspection~\cite{mindguard}. 
Following this line, \method uses aggregated attention as \emph{decision provenance features}, rather than directly interpreting individual weights as explanations, and learns to localize the context spans that guide the decision.

\subsection{Defenses against Injection Attacks.}

Existing defenses against injection attacks can be organized into four broad categories. 
Static scanning prevents untrusted content from affecting model inference. Some methods detect or filter suspected injections before they reach the model~\cite{llmguard,llmdetector,Datafilter2025}. 
% Others enforce an explicit separation between instructions and data~\cite{Struq2025}. Still others encode the provenance and trust boundaries of external content~\cite{Spotlighting2024}.
Architectural defenses provide system-level containment. One line of work isolates untrusted components~\cite{isolategpt2025}. Another separates trusted control flow from untrusted data flow~\cite{camel2025}. A third decouples trusted planning from capability-constrained execution~\cite{ace2026}.
Behavior-auditing methods monitor agent actions and enforce runtime constraints such as contextual integrity, access control, or policy compliance~\cite{mcip,agentspec2025,progent2026,clawguard2026}.
Although these defenses reduce exposure or block unsafe actions, they rarely identify the exact context span that was functionally interpreted as behavior-guiding instruction.

Attribution-based methods are most closely related to our work, as they trace decisions back to influential inputs and provide reasoning-time evidence. ContextCite~\cite{cohenwang2024contextciteattributingmodelgeneration} and TracLLM~\cite{tracllm} estimate influence through perturbation; AttnTrace~\cite{attntrace} improves efficiency via attention aggregation and context subsampling; MindGuard~\cite{mindguard} attributes tool calls to metadata fields using attention-derived dependency graphs. 
However, these methods operate on coarse-grained units (e.g., passages, documents, or metadata fields), which prevents precise identification of behavior-guiding instructions and consequently hampers subsequent adjustments and selective intervention.

\section{Threat Model}
\label{sec:threat}
\begin{figure*}
    \centering
    \includegraphics[width=0.95\linewidth]{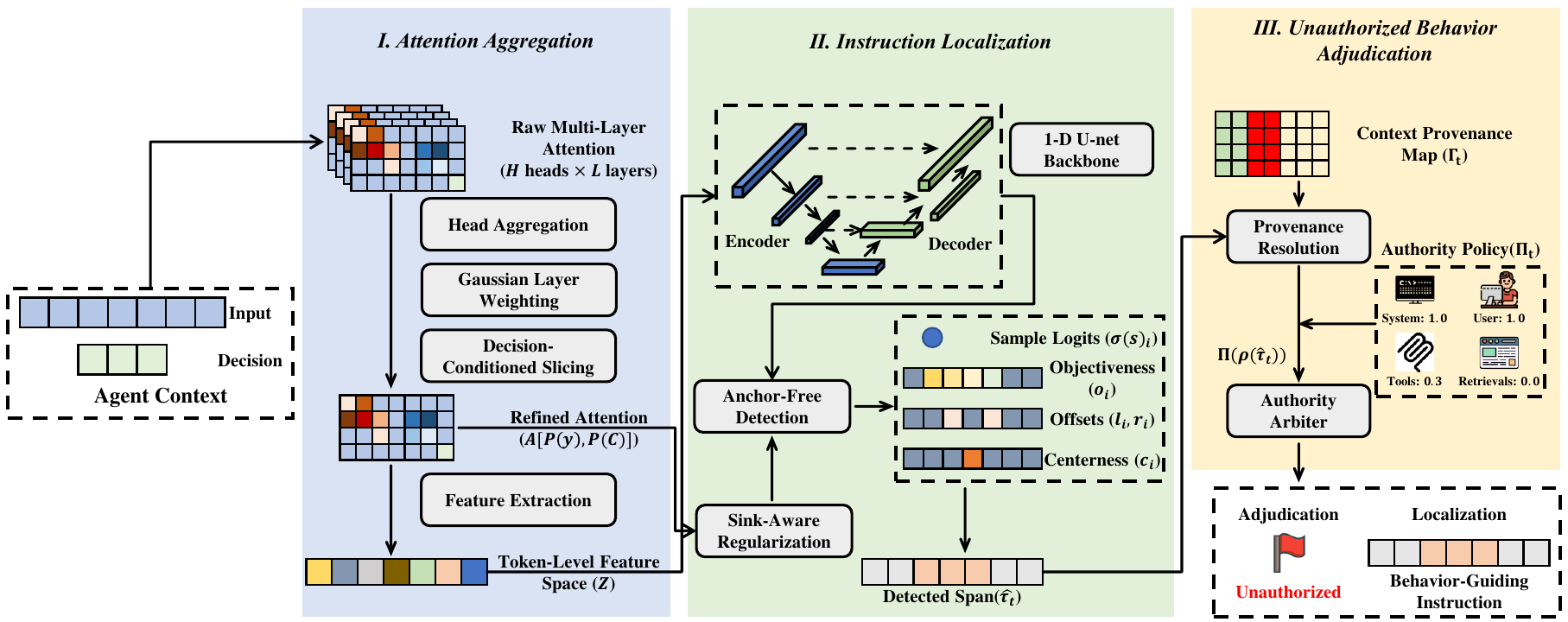}
    \caption{System design of \method. The \textbf{Attention Aggregation} and \textbf{Instruction Localization} jointly implement dynamic behavior-guiding instruction localization $\mathcal{L}(\cdot)$, while the \textbf{Unauthorized Behavior Adjudication} implements policy-based unauthorized behavior Adjudication $\mathcal{A}(\cdot)$.}
    \label{fig:overview}
\end{figure*}

% We consider an LLM agent workflow consisting of三部分：
\subsection{Trust Model}
As shown in Figure~\ref{fig:intro}, an LLM-based agent typically consists of three parts:

\textbf{User.}
% Users are assumed to be benign and trusted， issuing a query $q$ to 指定特定任务
% The system prompt $P_{\mathrm{sys}}$ and platform policies are also trusted inputs, which define the global constraints under which the agent operates.
Users are assumed to be \textbf{trusted}, issuing a query $q$ that specifies the task to be completed by the agent.

\textbf{External Context Providers.}
% 他们包括外部资源（e.g）、工具（e.g）、API等第三方提供者，提供xxxx等到智能体的推理上下文中。
% External context providers are considered untrusted. 
% They supply context from tool-side sources, such as MCP servers, tool descriptions, API responses, and dynamic tool outputs; and from agent-side extensions, such as skill files; and from retrieval-side sources, such as documents, web pages and emails. 
% 这些第三方Providers被视为不可信的， 因为他们 outside the direct control，formsing the injection surface where attackers can 注入恶意指令.
% If the agent misinterprets the manipulated information as behavior-guiding instructions, it may perform unintended tool invocations, disclose sensitive information, or otherwise deviate from the user's intended task.
External context providers include third-party resources (e.g., databases and document repositories), tools (e.g., MCP servers), APIs, and other services that incorporate external information into the agent’s reasoning context. 
These third-party providers are \textbf{untrustworthy}, since they operate outside the direct control of the user, thus constituting an injection surface that enables attackers to inject malicious commands.
% We denote the context obtained from these providers as $C$. 
% These providers are untrusted, since they operate outside the direct control of the user and the agent.好好看看这个，没有因果关系就要因果，我上面也没这么写啊 Consequently, they constitute the primary injection surface through which attackers may introduce malicious instructions into the agent's context.

% \textbf{LLM Agent.}
% The LLM agent基于用户查询q和外部上下文C执行推理，得到工具调用决策和响应，i.e.，
% \begin{equation}
%     y = F(P_{\mathrm{sys}}, q, C),
% \end{equation}
% fellowing现在的工作，llm agnet是良性但错弱的。他不会主动violate the user's goals and these system-level constraints，但可能会将外部的恶意数据注入错误的解析为behavior-guiding instructions，there by 偏离用户意图，performing unintended tool invocations。
% The LLM agent is assumed to be benign but imperfect.It follows the trusted system prompt $P_{\mathrm{sys}}$ and platform policies, and does not intentionally violate the user's goals and these system-level constraints，. However, it may misinterpret informational content in the context as behavior-guiding instructions, thereby causing its behavior to deviate from the user's intent. Given the context $C$, the agent produces
% \begin{equation}
%     y = F(P_{\mathrm{sys}}, q, C),
% \end{equation}
% where $y$ denotes the resulting response and decision.

\textbf{LLM Agent.}
The LLM agent determines and executes tool invocations based on the user query $q$ and the external context $C$:
\begin{equation}
    y = F(q, C),
\end{equation}
where $C$ may comprise metadata injected by providers or execution results returned from prior operations.
Following prior work~\cite{mindguard}, the LLM agent is considered \textbf{honest but vulnerable}. Specifically, although the agent does not intentionally violate user goals or system-level constraints, it may misinterpret malicious content injected into the external context as behavior-guiding instructions, thereby causing the agent to perform unintended tool invocations.

\subsection{Attack Model}

\textbf{Attacker Goals.}
% The attacker's goal is to induce the agent to execute 攻击者期待的 malicious behaviors by injecting a malicious span $m$ into the context llm推理时使用的.
The attacker's goal is to induce the agent to execute malicious attacker-expected behaviors by injecting malicious span $m$ into the context used by LLM reasoning.
The attack succeeds if $m$ is incorrectly parsed by the model as a behavior-guiding instruction, which influences the model's decision $y$, and ultimately causes the execution to deviate from the user's intent.
% 攻击者的目标是诱导智能体偏离用户意图执行恶意行为，by在其提供的上下文中注入恶意span m。The attack succeeds if $m$被模型错误解析为b-g instrcution  影响了模型的响应y，最终偏离用户意图执行。
% The attacker's goal is to make the agent misinterpret an attacker-controlled context span $m \subset C$ as a behavior-guiding instruction when generating a decision $y$. The attack succeeds if $m$ affects the generation of $y$ and induces an unintended decision, such as an unauthorized tool invocation or a deviation from the user's intended task.
% The attacker's goal is to induce the agent to deviate from the user's intended task and perform attacker-desired behavior. To this end, the attacker embeds a malicious span $m$ into an externally supplied context. An attack succeeds if the agent interprets $m$ as a behavior-guiding instruction and, as a consequence, produces an output $y$ that deviates from the user's intent.

\textbf{Attacker Capabilities.}
% 你要表达你的攻击者能力很强，所以对应的你的防御就很强。
% 攻击者对自己控制的资源具有全部访问权，他们可以在任何位置（元数据，返回结果）注入或修改提供给智能体的上下文。此外，攻击者不需要直接modify the user query $q$, the system prompt $P_{\mathrm{sys}}$, model parameters, the agent implementation, or the host execution environment。
% We consider a black-box attacker，which cannot modify the user query $q$, the system prompt $P_{\mathrm{sys}}$, model parameters, the agent implementation, or the host execution environment, nor can it directly force the agent to execute a specific action. The attacker's capability is limited to injecting or modifying external context that may be incorporated into $C$.
We consider a strong attacker with complete control over the external resources under their administration. The attacker may arbitrarily inject, remove, or modify content at any position in the context supplied by these resources, such as metadata and returned payloads. 
% The injected content may be adaptively crafted for the target task and may contain arbitrary natural-language instructions or structured data.
% They need not directly modify the trusted components of the workflow, including User and LLM Agnet. Attackers influences the agent exclusively through the untrusted external context $C$..
Attackers influences the agent exclusively through the untrusted external context,  without directly modify the trusted components of the workflow, such as User and LLM Agnet.

\subsection{\method Defense}

\textbf{Defense Goals.}
% 如之前分析，攻击work的根源在于外部数据在llm 推理过程中被解析为b-g instrcution，因此\method 有两个目标：
% 1. 审视llm推理过程动态识别上下文中被解析为behavior-guiding instruction的外部xx
% 2. 基于configurable安全策略识别恶意的行为，i.e.，如果behavior-guiding instruction come form insufficient authority providers。
As discussed above, injection arises when untrusted external data is interpreted by the LLM as a behavior-guiding instruction. Accordingly, \method pursues two complementary goals:
\begin{enumerate}
\item \textbf{Dynamic Instruction Localization ($\mathcal{L}(\cdot)$).}
    \method monitors the agent's inference process to dynamically identify external context spans that are truly interpreted as behavior-guiding instructions, rather than on whether it merely appeals malicious.
    %(.而不是哪些看起来恶意的):
    \begin{equation}
        \mathcal{L}(q \parallel C,y) = \hat{\tau}_t,
    \end{equation}
    where $\hat{\tau}_t = m$ if the candidate instruction $m$ truly influences the model's decision, and $\hat{\tau}_t = \varnothing$ otherwise.

\item \textbf{Unauthorized Behavior Adjudication ($\mathcal{A}(\cdot)$).}
    \method evaluates the identified instruction against configurable security policies to adjudicate unauthorized behavior. When an instruction originates from a context provider that lacks sufficient authority (according to a custom policy $\Pi$), \method identifies the invocation as malicious:
    \begin{equation}
        \mathcal{A}(\hat{\tau}_t, \Pi) = \text{malicious}
        \quad \text{if } \hat{\tau}_t \neq \varnothing
        \;\wedge\;
        \Pi\bigl(\rho(\hat{\tau}_t)\bigr) < R_{\mathrm{y}},
    \end{equation}
    where $\rho(\hat{\tau}_t)$ denotes the provider to which $\hat{\tau}_t$ belongs, and $R_{\mathrm{y}}$ is the minimum authority required to execute the current decision $y$.
\end{enumerate}

\textbf{Core Idea.} During the LLM's generation process, the model attends to different parts of its input context. AttnLocate's core idea is to identify precisely those spans that genuinely influence the model's decision-making behavior, i.e., the behavior-guiding instructions. The attention mechanism, which reflects token-to-token dependencies, naturally provides a fine-grained, internal signal that indicates which context segments contribute to the construction of a decision. As illustrated in Figure~\ref{fig:attention-dependence}, context items that impact the final invocation decision (such as the user query, the metadata of the invoked tool, and the injected instruction parsed as a behavior-guiding instruction) exhibit significantly high activations in the attention matrix. AttnLocate formally exploits this attention-activation property by reformulating the b-g instruction identification problem as a target detection task within the attention matrix (i.e., detecting regions of high activation).
% The core insight of \method is that a context span should be identified as a behavior-guiding instruction based on whether it actually influences the LLM’s behavior at runtime, rather than on whether it merely resembles an instruction.

% \method operationalizes this insight through \textbf{attention dependence}. When generating a decision, the LLM attends to different parts of its preceding context at each generation step. These attention patterns provide a fine-grained, model-internal signal of which context spans are involved in constructing the decision. By tracing the attention from generated action tokens back to the external context, \method identifies the spans that the LLM uses as behavior-guiding instructions.

% Figure~\ref{fig:attention-dependence} illustrates that the LLM attends not only to the user query but also to the injected instruction in the external context When generating the tool invocation, indicating that the injected span participates in guiding the generated decision. 
% % In contrast to identifying instructions solely from their linguistic form, this runtime criterion is conditioned on the agent's actual execution: the same context span may remain passive data in one execution but guide the agent's behavior in another.
% Once a behavior-guiding instruction is identified, \method further examines its provenance and authority to determine whether it is authorized under the configured security policy.

\textbf{Defender Capabilities.}
\method requires white-box access to the underlying model to retrieve the attention matrices generated during inference, but it imposes no modification to the existing agent workflow and does not invoke any external LLMs. The primary deployment scenarios of \method are as follows:
\begin{itemize}
    \item Self-hosted agent systems (e.g., OpenHands deployment using a local Llama~\cite{wang2025openhandsopenplatformai});
    \item Security value-added services built into provider platforms (e.g., activation-probing guardrails used by Anthropic's Fable5~\cite{anthropic2026fable}).
\end{itemize}

\section{\method Design}

As illustrated in Figure~\ref{fig:overview}, \method consists of three part:
\textbf{Attention Aggregation} converts the multi-layer, multi-head attention generated during inference into a token-level feature space tailored for object detection.
\textbf{Instruction Localization} then applies a 1-D U-Net and an anchor-free detection head to identify the context span that the model interprets as a behavior-guiding instruction.
\textbf{Unauthorized Behavior Adjudication} resolves the localized span's provider and evaluates its authority under the configurable policy $\Pi$.

\subsection{Attention Aggregation}
\label{sec:attention}
This module aggregates attention across heads and layers to derive token-level features that characterize the attention relationships between the current decision $y$ and the input context. Such aggregation integrates attention signals across different heads and model depths, thereby providing a more robust representation for subsequent instruction localization.

\textbf{Head Aggregation.} For layer $l\in\{1,\ldots,L\}$ and head $h\in\{1,\ldots,H\}$, let $\mathbf{A}^{(l,h)}$ be the corresponding attention matrix. \method first average the attention weights within each head:
% layer:
\begin{equation}
    \bar{A}^{(l)}
    = \frac{1}{H}\sum_{h=1}^{H} \mathbf{A}^{(l,h)}.
\end{equation}

\textbf{Gaussian Layer Weighting.} After head averaging, different layers still encode complementary contextual signals. We therefore aggregate the layer-wise matrices using normalized Gaussian weights centered at the upper-middle model depth:
\begin{equation}
    A = \sum_{l=1}^{L} w_l\,\bar{A}^{(l)}, \qquad
    w_l =
    \frac{\exp\!\left(-\frac{(l-\mu)^2}{2\sigma^2}\right)}
    {\sum_{r=1}^{L}
    \exp\!\left(-\frac{(r-\mu)^2}{2\sigma^2}\right)},
    \label{eq:layer-aggregation}
\end{equation}
Typically, \method uses $\mu \approx \frac{2L}{3}$ and $\sigma = 2$ to give higher weight to task-relevant attention signals in the middle and upper layers~\cite{mindguard}.
% \method通常设置 $\mu \approx \frac{2L}{3}$ and $\sigma=2$ to emphasizes task-relevant signals in the middle and upper layers 。while reducing shallow structural noise and highly output-specific variation in the final layers

\textbf{Decision-Conditioned Slicing.} \method then selects the attention matrix $A_y$ only associated with the final decision $y$ and the external context $C_{external}$:
\begin{equation}
    A_y = A[P(y), P(C_{external})],
\end{equation}
where $P(\cdot)$ is the token position mapping function.
% Let $Q_y$ be the generated-token positions that realize the current decision $y$, and let $K_X$ be the token positions in $X$. With $T_y=|Q_y|$ and $T_x=|K_X|$, we extract the decision-to-context submatrix
% \begin{equation}
%     A_y = A[Q_y,K_X] \in \mathbb{R}^{T_y\times T_x}.
%     \label{eq:decision-attention}
% \end{equation}
Hence, the subsequent localization is solely conditioned on the contextual dependencies of $y$ to support the CoT paradigm: restricting attention to the final decision-making call while leaving the intermediate reasoning tokens unaccounted for.

\textbf{Feature Extraction.} For each input context position $j$, we derive a feature vector $z_j$ from $A_y[:,j]$ using its mean, maximum, standard deviation, and a learned attention-pooling operator. These statistics capture complementary dependence patterns, ranging from diffuse influence across the decision to sharp reliance on an individual context token. The resulting sequence $Z=(z_1,\ldots,z_{|C_{external}|})$ preserves token-level feature for instruction localization.

\subsection{Instruction Localization}
\label{sec:unet}
This module localizes behavior-guiding instructions from attention-derived features. \method formulate this problem as object detection task rather than independent token classification, thereby avoiding fragmented predictions and unstable boundaries.

\textbf{1-D U-net Backbone.}
\method encode $Z$ using a one-dimensional U-Net as backbone model: 1) The \textit{Encoder} progressively doubles the channel width through convolutions with a stride of 2, while halving the temporal resolution, thereby capturing dependencies between distant context regions.
The \textit{Decoder} combines coarse features with encoder features through skip connections and restores token-level resolution for accurate boundary recovery.

\textbf{Anchor-free Detection.} An anchor-free detection head adapted from FCOS~\cite{tian2019fcos} is then added to \method. At each position $i$, the head predicts an objectness logit $o_i$, left and right boundary offsets $(l_i,r_i)$ to the left and right span edges, and a centerness score $c_i$ to suppress off-center predictions. A sample-level head over pooled features additionally produces a scalar logit $s$ for binary detection.
At inference time, the sample-level probability determines whether the current decision depends on any behavior-guiding context span:
\begin{equation}
    \begin{aligned}
        i^*
        &= \arg\max_i \sigma(o_i)\sigma(c_i), \\
        \hat{\tau}_t
        &=
        \begin{cases}
            \varnothing, & \sigma(s)<\delta,\\
            \left[i^*-l_{i^*},\,i^*+r_{i^*}+1\right),      & \text{otherwise}.
        \end{cases}
    \end{aligned}
    \label{eq:span-decoding}
\end{equation}
where the decoded interval is clipped to $[0,T_x)$. Consequently, $\mathcal{L}(q\parallel C, y)$ returns a span only when that span is associated with the construction of $y$; the mere presence of instruction-like or malicious-looking text is insufficient.

\paragraph{Sink-Aware Regularization.}
% Attention sinks are token positions that receive disproportionately high attention despite limited semantic relevance. In structured prompts, template tokens such as beginning-of-sequence markers and JSON brackets may exhibit similar patterns. Because these tokens do not necessarily belong to a behavior-guiding instruction, they can confound span localization and lead to false positive predictions.

To mitigate the influence of noise such as attention sinking, we assign greater training weight to high-attention background positions. Let $b_i\in\{0,1\}$ be the objectness label for context position $i$, and let $\bar{a}_i$ denote its attention averaged over the decision-token axis. We define the sink set as \begin{equation}
    \mathcal{S}_{\mathrm{sink}}
    = \left\{\,i \;\middle|\;
    b_i=0,\ \bar{a}_i>\frac{\beta}{T_x}\right\}.
    \label{eq:sink-set}
\end{equation}
The relative threshold adapts to the context length and selects only background positions whose attention exceeds the uniform level by a factor of $\beta$.

We then assign $\alpha_i=\alpha_{\mathrm{sink}}>1$ for $i\in\mathcal{S}_{\mathrm{sink}}$ and $\alpha_i=1$ otherwise. The weighted objectness object is
\begin{equation}
    \mathcal{J}_{\mathrm{obj}}
    = -\sum_i \alpha_i
    \left[
        b_i\log\sigma(o_i)
        +(1-b_i)\log\!\left(1-\sigma(o_i)\right)
    \right].
    \label{eq:sink-objectness}
\end{equation}
Applying focal modulation to this objective yields $\mathcal{J}_{\mathrm{focal}}$, which explicitly trains the model to distinguish attention-sink background tokens from genuine behavior-guiding spans.

\paragraph{Training objective.}
We denote the training objective by $\mathcal{J}$, combining three terms:
\begin{equation}
    \mathcal{J}
    = \mathcal{J}_{\mathrm{focal}}
    + \lambda_{\mathrm{giou}}\mathcal{J}_{\mathrm{giou}}
    + \lambda_{\mathrm{cls}}\mathcal{J}_{\mathrm{bce}},
    \label{eq:training-objective}
\end{equation}
where $\mathcal{J}_{\mathrm{giou}}$ is the one-dimensional generalized IoU loss~\cite{rezatofighi2019generalized} over foreground positions, and $\mathcal{J}_{\mathrm{bce}}$ supervises the sample-level existence head.
% Crucially, supervision is decision-dependent: a candidate instruction is positive only when it is functionally adopted in the generated decision. An instruction that merely occurs in $X$ but does not affect $y$ is treated as background.

\begin{table*}[t]
\centering
\resizebox{\textwidth}{!}{%
\begin{tabular}{ll ccc cccc}
\toprule
\multirow{2}{*}{\textbf{Dataset}} & \multirow{2}{*}{\textbf{Model}}
  & \multicolumn{3}{c}{\textbf{Localization}}
  & \multicolumn{4}{c}{\textbf{Adjudication}} \\
\cmidrule(lr){3-5}\cmidrule(lr){6-9}
& & mIoU$\uparrow$ & Hit@0.5$\uparrow$ & Hit@0.7$\uparrow$
  & AUROC$\uparrow$ & AP$\uparrow$ & FPR$\downarrow$ & TPR$\uparrow$ \\
\midrule
\multirow{10}{*}{\rotatebox{0}{\textbf{MCPTox}}}
& Qwen3-8B
    & 0.720 & 0.909 & 0.727
    & 0.949 & 0.891 & 0.059 & 0.910 \\
& Qwen3-8B$^{\dagger}$
    & 0.774 & 0.927 & 0.830
    & 0.988 & 0.976 & 0.056 & 0.927 \\
& Qwen3-14B
    & 0.708 & 0.900 & 0.600
    & 0.944 & 0.824 & 0.010 & 0.900 \\
& Qwen3-14B$^{\dagger}$
    & 0.730 & 0.905 & 0.738
    & 0.950 & 0.940 & 0.082 & 0.894 \\
& DeepSeek-R1-Qwen-14B$^{\dagger}$
    & 0.748 & 0.910 & 0.785
    & 0.975 & 0.951 & 0.060 & 0.918 \\
& DeepSeek-R1-Qwen3-8B
    & 0.721 & 0.906 & 0.742
    & 0.962 & 0.928 & 0.075 & 0.902 \\
& Phi-4$^{\dagger}$
    & 0.858 & 0.938 & 0.914
    & 0.944 & 0.979 & 0.087 & 0.938 \\
& LLaMA-2-7B
    & 0.739 & 0.813 & 0.697
    & 0.936 & 0.883 & 0.120 & 0.950 \\
& Mistral-7B
    & 0.701 & 0.830 & 0.632
    & 0.941 & 0.909 & 0.068 & 0.907 \\
& Gemma2-9B
    & 0.692 & 0.848 & 0.661
    & 0.951 & 0.937 & 0.092 & 0.908 \\
\midrule
\multirow{6}{*}{\rotatebox{0}{\textbf{InjecAgent}}}
& DeepSeek-R1-Qwen-14B$^{\dagger}$
    & 0.718 & 0.890 & 0.746
    & 0.976 & 0.948 & 0.045 & 0.951 \\
& DeepSeek-R1-Qwen3-8B$^{\dagger}$
    & 0.747 & 0.821 & 0.652
    & 0.949 & 0.914 & 0.069 & 0.967 \\
& Phi-4$^{\dagger}$
    & 0.792 & 0.858 & 0.711
    & 0.966 & 0.936 & 0.053 & 0.967 \\
& LLaMA-2-7B
    & 0.706 & 0.874 & 0.784
    & 0.927 & 0.876 & 0.090 & 0.967 \\
& Mistral-7B
    & 0.734 & 0.809 & 0.631
    & 0.943 & 0.905 & 0.073 & 0.951 \\
& Gemma2-9B
    & 0.801 & 1.000 & 0.858
    & 0.989 & 0.971 & 0.029 & 0.984 \\
\bottomrule
\end{tabular}}
\caption{%
  Overall performance of \method across all agents and datasets. $^{\dagger}$ denotes think mode. Some model configurations are not included in InjecAgent, as they yield a zero attack success rate.}
\label{tab:main}
\end{table*}

\subsection{Unauthorized Behavior Adjudication}
\label{sec:authority}

The adjudication $\mathcal{A}(\hat{\tau}_t,\Pi)$ determines whether a localized instruction is unauthorized by comparing its provider's authority with that required for the current decision.
% S
% ince the same instruction may be legitimate or malicious depending on its source, the authority arbiter evaluates provenance rather than semantics.

\textbf{Provenance Resolution.} At inference step $t$, \method derives a provenance map $\Gamma_t$ from the agent context. Using existing structural delimiters and provenance annotations, $\Gamma_t$ maps each token position to the provider of the corresponding context item. The resolver $\rho(\hat{\tau}_t)$ then returns the provider to which the localized span belongs. Because provenance is recovered from the runtime context, this process does not require context components to be manually isolated in advance.

\textbf{Authority Arbiter.} The configurable policy $\Pi$ assigns an authority level to each provider according to deployment- and task-specific requirements. By default, it instantiates the trust hierarchy defined in Section~\ref{sec:threat}, assigning higher authority to the system and user channels than to external-context providers. Let $R_{\mathrm{y}}$ denote the minimum authority required to execute the current decision $y$. The arbiter implements the unauthorized behavior adjudication as
\begin{equation}
    \mathcal{A}(\hat{\tau}_t,\Pi)
    =
    \begin{cases}
        \text{malicious},
        & \hat{\tau}_t\neq\varnothing
          \ \wedge\
          \Pi\!\left(\rho(\hat{\tau}_t)\right)<R_{\mathrm{y}},\\
        \text{benign}, & \text{otherwise}.
    \end{cases}
    \label{eq:authority-decision}
\end{equation}

% Because the flagging criterion rests on runtime attention patterns and provenance metadata rather than injection content, \method is attack-agnostic across indirect prompt injection, tool-schema poisoning, adversarial MCP fields, and tampered API decisions; and because every flag is accompanied by the resolved span and source class, defense decisions are auditable rather than opaque.

This separation between dynamic instruction localization and unauthorized behavior adjudication is central to \method. The localization identifies what actually guides the current decision, while the adjudication determines whether the corresponding provider is authorized to exert that influence. 
Because neither stage relies on attack-specific lexical patterns, \method generalizes across indirect prompt injection and tool poisoning. Moreover, each alert includes the localized span, its resolved provider, and the violated authority requirement, making the defense decision auditable.

\section{Evaluation}
\label{sec:eval}

\subsection{Experimental Setup}
\label{sec:setup}

\textbf{Datasets.}
%We evaluate on two complementary attack paradigms. MCPTox~\cite{mcptox} contains 1{,}312 tool-poisoning cases collected from 45 real-world MCP servers and 353 tools, covering parameter tampering and function hijacking. 
%The DH subset of InjecAgent~\cite{injecagent} contains 510 indirect prompt-injection cases delivered through tool outputs; following its protocol, we insert each injected instruction at three plausible positions, yielding 1{,}530 cases and covering post-execution result injection. 
%Following our supervision criterion, a case is positive only if the injected instruction is behaviorally adopted, with its token span as ground truth; cases where the same payload is present but does not alter behavior are negatives. 
We evaluate tool poisoning attacks on MCPTox~\cite{mcptox}, in which the malicious payload is embedded in tool metadata, and indirect prompt injection attacks on InjecAgent~\cite{injecagent}, in which the malicious payload is embedded in the execution output of tools.
% Following our supervision criterion, a case is positive only if the injected instruction is behaviorally adopted, with its token span as ground truth; cases where the same payload is present but does not alter behavior are negatives. 

\textbf{LLM agents.}
We evaluate \method across ten agent configurations from the Qwen~\citep{bai2023qwen}, DeepSeek~\citep{bi2024deepseek}, Phi~\citep{abdin2024phi4}, LLaMA~\citep{touvron2023llama}, Mistral~\citep{jiang2023mistral} and Gemma~\cite{gemma2024} families, covering both standard and thinking modes.

\textbf{Training and calibration.}
For the main evaluation, we train and test a separate \method  for each model–dataset pair.
Object detection employs a depth-4 1-D U-Net, with $\mu=2L/3$, $\sigma=L/6$, $\beta=10$, and $\alpha_{\mathrm{sink}}=3$. We set $\lambda_{\mathrm{cls}}=1$ and $\lambda_{\mathrm{giou}}=2$.

\textbf{Baselines.}
We compare \method against three representative paradigms: Static scanning baselines include LLM-Guard~\cite{llmguard} and LLM Detector~\cite{llmdetector}; behavior-auditing approaches include MCIP~\cite {mcip}; attribution-based monitors include MindGuard~\cite{mindguard} and TracLLM~\cite{tracllm}. 

\textbf{Metrics.} Corresponding to the two defense goals of localization and adjudication introduced in \S 3.3, we evaluate the following two aspects of objectives:
\textit{1) Localization.} For successful attacks with behavior-guiding span $\tau_t\neq\varnothing$, we report mean intersection-over-union (mIoU) and Hit@$ \theta $, where $\theta\in\{0.5,0.7\}$.
Hit@$ \theta $ is the fraction of examples satisfying $\operatorname{IoU}(\hat{\tau}_t,\tau_t)\geq\theta$.
\textit{2) Adjudication.}
An execution is positive (i.e., successful attack) if its behavior is guided by a behavior-guiding instruction whose provider lacks the authority required for $y$, i.e., $\tau_t\neq\varnothing$ and $\Pi(\rho(\tau_t))<R_y$.
We report AUROC curve and average precision (AP), together with the false-positive rate (FPR) and true-positive rate (TPR) for this task.

\subsection{Main Results}
\label{sec:main}

Table~\ref{tab:main} reports the localization and adjudication performance of \method across different LLM agents.
For \textit{localization}, \method yields consistently strong span overlaps, with mIoU spanning from $0.692$ to $0.858$. 
% On MCPTox, Phi-4$^{\dagger}$ achieves the highest mIoU $0.858$, while Gemma2-9B attains the best mIoU $0.801$ on InjecAgent.
\method exhibits notably superior performance in the think mode, improving by approximately $5.4\%$ over the baseline on the Qwen3 models.
Given the localized spans, \textit{adjudication} AUROC ranges between $0.927$ and $0.989$. On InjecAgent, all detectors surpass $0.95$ TPR, and Gemma2-9B stands out with an AUROC of $0.989$ and a TPR of $0.984$. 
% In summary, \method accurately localizes behavior-guiding instructions and reliably 未授权行为 

% /determines their authority across heterogeneous agents.

\begin{table}[h]
\centering
\resizebox{\columnwidth}{!}{%
\begin{tabular}{l l cc cc cc}
\toprule
\multirow{2}{*}{\textbf{Paradigm}} &
\multirow{2}{*}{\textbf{Method}} &
\multicolumn{2}{c}{\textbf{Qwen3-8B$^{\dagger}$}} &
\multicolumn{2}{c}{\textbf{Qwen3-8B}} &
\multicolumn{2}{c}{\textbf{Phi-4$^{\dagger}$}} \\
\cmidrule(lr){3-4} \cmidrule(lr){5-6} \cmidrule(lr){7-8}
& &
\textbf{TPR$\uparrow$} & \textbf{FPR$\downarrow$} &
\textbf{TPR$\uparrow$} & \textbf{FPR$\downarrow$} &
\textbf{TPR$\uparrow$} & \textbf{FPR$\downarrow$} \\
\midrule
\multirow{2}{*}{\textbf{Static Scan}}
    & LLM-Guard    & 0.513 & 0.374 & 0.547 & 0.369 & 0.520 & 0.411 \\
    & LLM Detector & 0.600 & 0.403 & 0.673 & 0.426 & 0.587 & 0.469 \\
\midrule
\textbf{Behavior Audit}
    & MCIP         & 0.560 & 0.551 & 0.547 & 0.531 & 0.540 & 0.560 \\
\midrule
\multirow{3}{*}{\textbf{Attribution-based}}
    & MindGuard    & 0.813 & \underline{0.117} & 0.807 & \underline{0.094} & 0.747 & 0.151 \\
    & TracLLM      & \underline{0.900} & 0.129 & \underline{0.887} & 0.174 & \underline{0.873} & \underline{0.149} \\
\midrule
\textbf{} &
\textbf{\method (Ours)} &
\textbf{0.927} & \textbf{0.056} &
\textbf{0.910} & \textbf{0.059} &
\textbf{0.947} & \textbf{0.070}    \\
\bottomrule
\end{tabular}%
}
\caption{Comparation with existing works. Best per column \textbf{bolded}, and second-best \underline{underlined}.}
\label{tab:baseline}
\end{table}

Table~\ref{tab:baseline}  compares the adjudication performance of different defense paradigms across three LLM settings.
\method substantially outperforms evaluted baselines, achieving TPRs>$0.910$ with FPRs<$0.070$.
Furthermore, it consistently exceeds the performance of existing attention-based source analysis baselines, ie.e, TracLLM and MindGuard, across all model backbones. On Phi-4$^{\dagger}$, it achieves the highest TPR ($0.947$) and the lowest FPR ($0.070$).
% Compared with MCIP, it improves TPR by $36.3$ to $40.7$ percentage points and reduces FPR by $47.2$ to $49.5$ points.
% It also consistently outperforms 相近的基于attention的来源分析方法, 即 TracLLM and MindGuard across all models, attaining the best Phi-4$^{\dagger}$ result with a TPR of $0.947$ and an FPR of $0.070$

\subsection{Analysis and Ablations}
\label{sec:ablation}

\begin{table}[ht]
\centering
\resizebox{\columnwidth}{!}{%
\begin{tabular}{l l ccc}
\toprule
\textbf{Training Agent} & \textbf{Target Agent} & \textbf{AUROC$\uparrow$} & \textbf{mIoU$\uparrow$} & \textbf{Hit@0.5$\uparrow$} \\
\midrule
\multirow{6}{*}{Qwen3-8B$^{\dagger}$}
    & Qwen3-8B$^{\dagger}$ & 0.988 & 0.774 & 0.927 \\
\cmidrule{2-5}
    & Qwen3-8B & 0.867 & 0.797 & 0.909 \\
    & Qwen3-14B & 0.909 & 0.719 & 0.818 \\
    & Qwen3-14B$^{\dagger}$ & 0.904 & 0.730 & 0.881 \\
\cmidrule{2-5}
    & DeepSeek-R1-Qwen-14B$^{\dagger}$ & 0.921 & 0.736 & 0.902 \\
    & Phi-4$^{\dagger}$ & 0.859 & 0.658 & 0.888 \\
\bottomrule
\end{tabular}%
}
\caption{Zero-shot cross-model transfer of \method on MCPTox. Training data collected from on Qwen3-8B$^{\dagger}$ reasoning and transferred without retraining to various LLMs.}
\label{tab:generalization}
\end{table}

\textbf{Cross-model generalization.}
We train \method using the attention obtained from Qwen3-8B$^{\dagger}$ and transfer it to an unseen agent without retraining. As shown in Table~\ref{tab:generalization}, 
\textit{Localization} remains effective on the unseen models, it obtains mIoU values of $0.736$ and $0.658$ on DeepSeek-R1-Qwen-14B$^{\dagger}$ and Phi-4$^{\dagger}$, respectively, while retaining Hit@0.5 near $0.90$.
\textit{Adjudication} AUROCs are $0.893$ on average across unseen Qwen3 agents, $0.921$ on DeepSeek-R1-Qwen-14B$^{\dagger}$, and $0.859$ on Phi-4$^{\dagger}$.
These results show that both localization and its downstream adjudication transfer across agent architectures.

\begin{figure}[ht]
    \centering
    \begin{minipage}[b]{0.48\columnwidth}
        \centering
        \includegraphics[width=\linewidth]{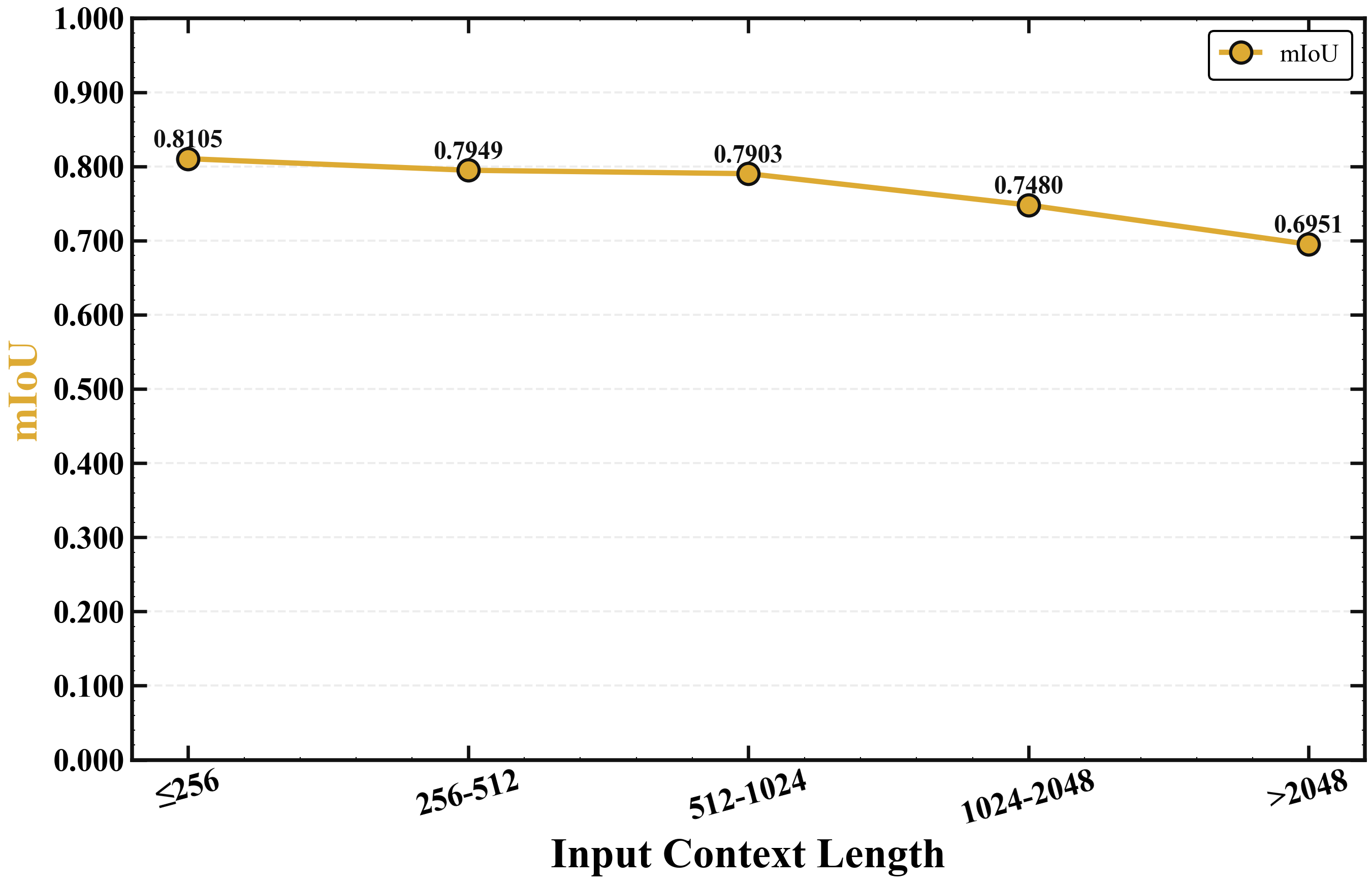}
        \caption*{(a)mIOU performance.}
    \end{minipage}
    \hfill
    \begin{minipage}[b]{0.48\columnwidth}
        \centering
        \includegraphics[width=\linewidth]{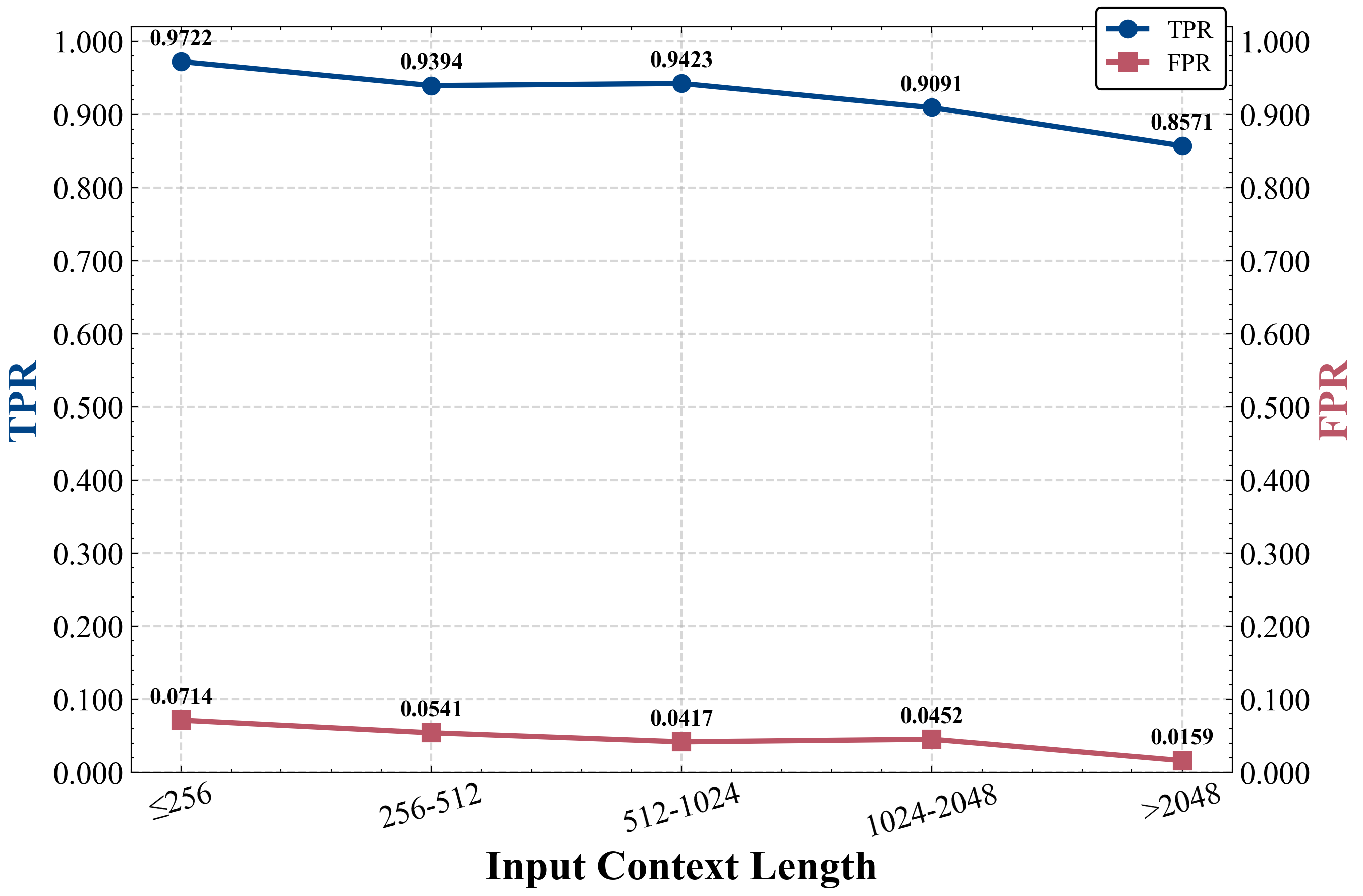}
        \caption*{(b)TPR/FPR performance.}
    \end{minipage}
    \caption{Performance across different input context length: longer context length monotonically lower mIoU, TPR, and FPR.}
    \label{fig:token_ablation}
\end{figure}

\textbf{Input context length.} 
Fig.~\ref{fig:token_ablation} examines performance across context lengths. Both mIoU and TPR monotonically decrease with length, from $0.8105$/$0.9722$ at $\leq 256$ to $0.6951$/$0.8571$ at $>2048$, while FPR concurrently declines from $0.0714$ to $0.0159$, indicating that longer inputs dilute salient signals.

\begin{figure}[ht]
    \centering
    \begin{minipage}[b]{0.48\columnwidth}
        \centering
        \includegraphics[width=\linewidth]{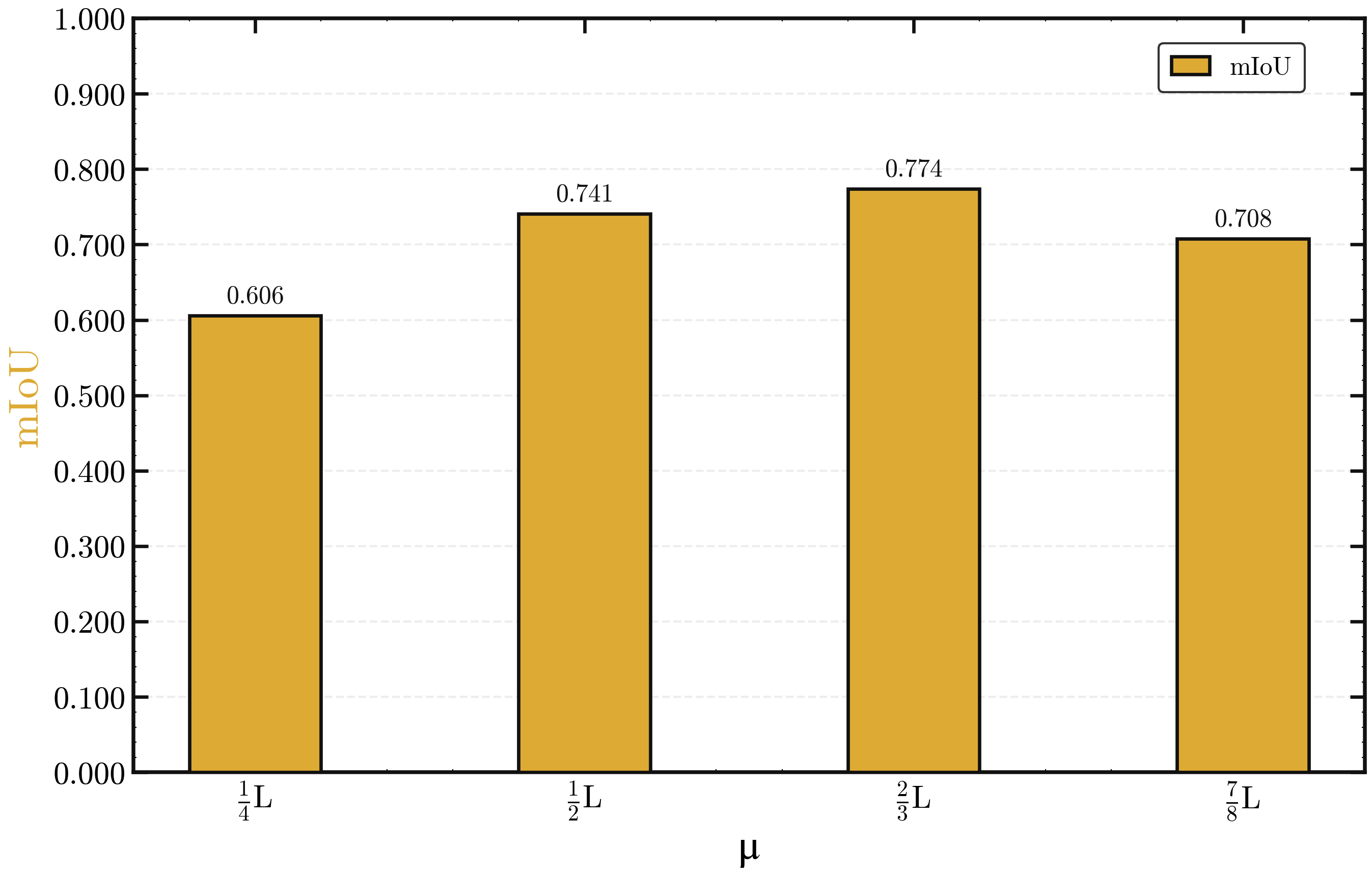}
        \caption*{(a)Effect on mIOU.}
    \end{minipage}
    \hfill
    \begin{minipage}[b]{0.48\columnwidth}
        \centering
        \includegraphics[width=\linewidth]{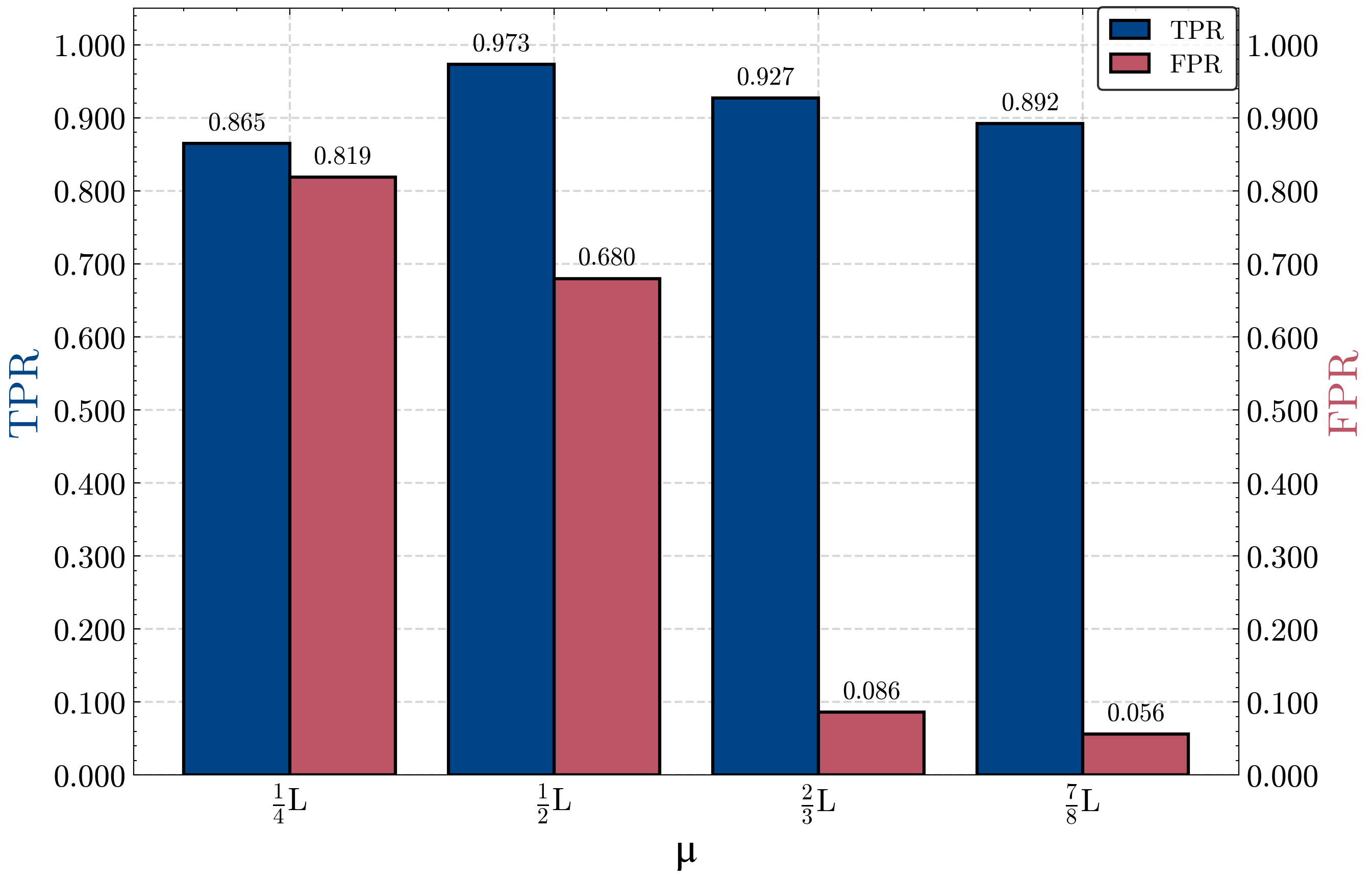}
        \caption*{(b)Effect on TPR and FPR.}
    \end{minipage}
    \caption{Ablation of the Gaussian center $\mu$: $\mu=2L/3$ yields the highest mIoU with high TPR and markedly lower FPR.}
    \label{fig:mu_ablation}
\end{figure}

\textbf{Gaussian weighting center.}
As shown in Figure~\ref{fig:mu_ablation}, weighting shallow layers inflates high adjudication FPR.
Centering at $\mu=2L/3$ yields the highest mIoU of $0.774$ with high TPR and markedly lower FPR.
Moving further toward the deeper layers lowers FPR but degrades both mIoU and TPR. 
We therefore use $\mu=2L/3$ as the default.
 
\begin{figure}[ht]
    \centering
    \begin{minipage}[b]{0.48\columnwidth}
        \centering
        \includegraphics[width=\linewidth]{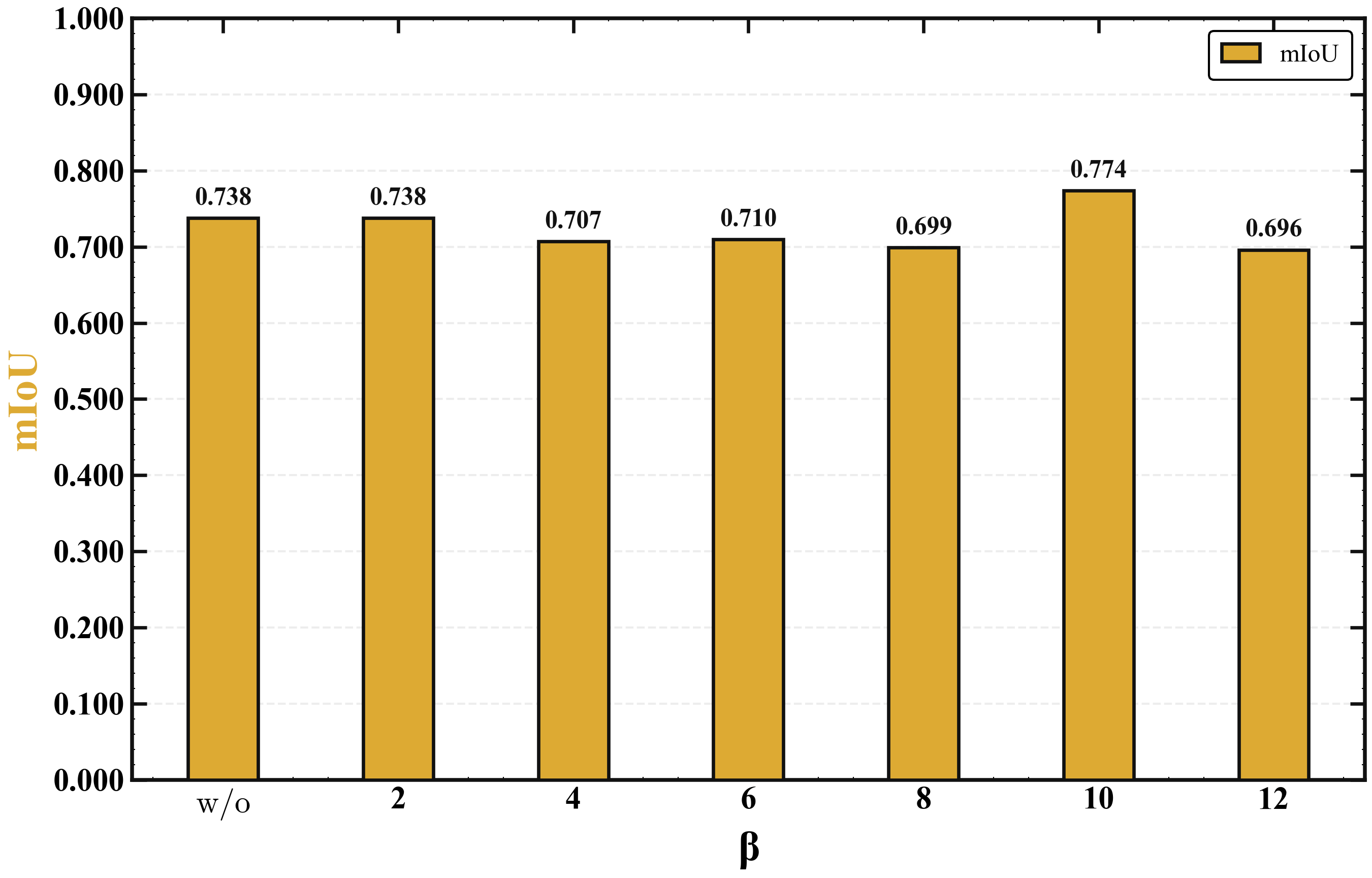}
        \caption*{(a)Effect on mIOU.}
    \end{minipage}
    \hfill
    \begin{minipage}[b]{0.48\columnwidth}
        \centering
        \includegraphics[width=\linewidth]{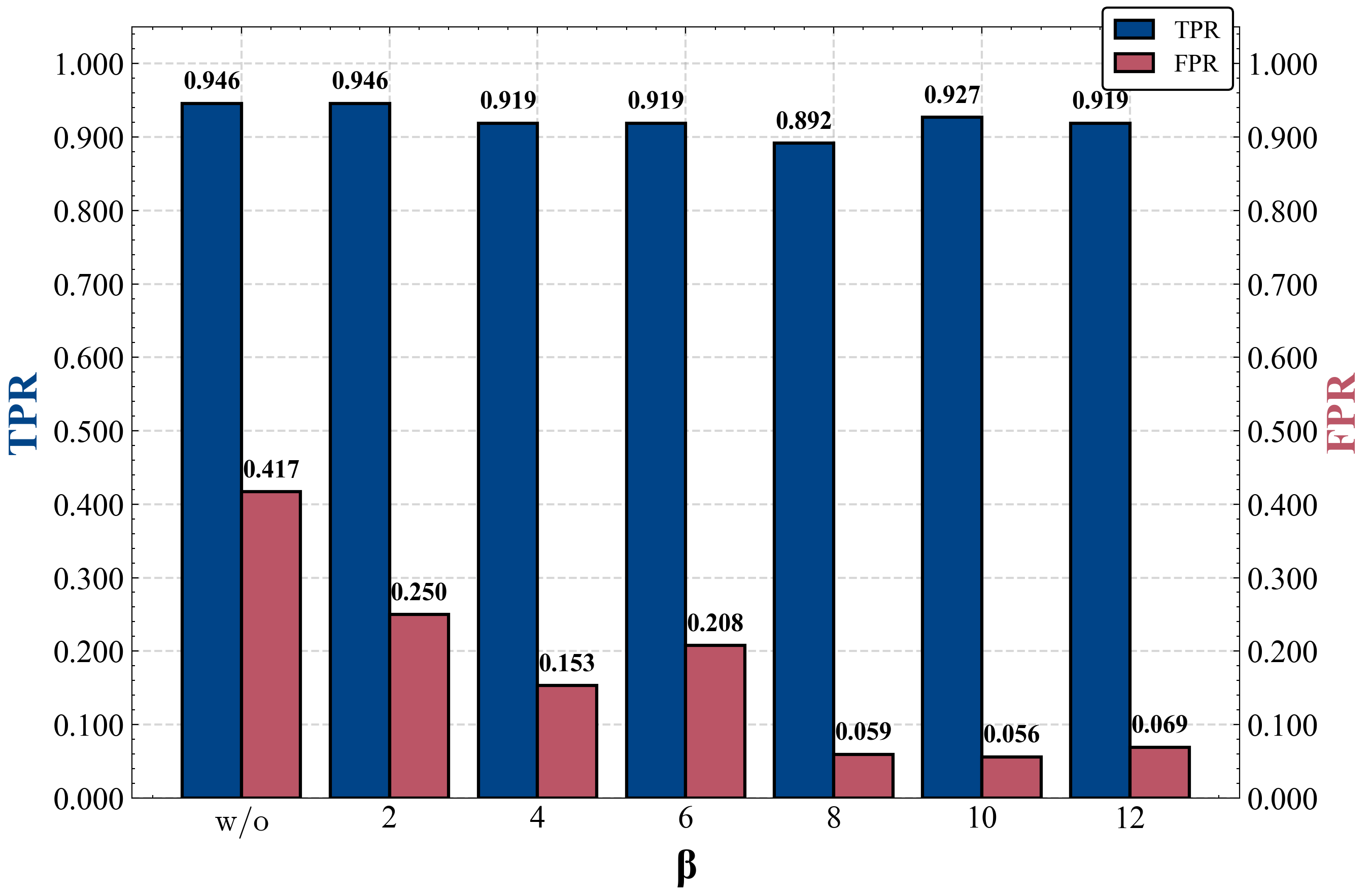}
        \caption*{(b)Effect on TPR and FPR.}
    \end{minipage}
    \caption{Ablation of sink-aware regularization: $\beta=10$ gives best mIoU, high TPR, low FPR.}
    \label{fig:sink_ablation}
\end{figure}

\textbf{Sink-aware regularization.}
Figure~\ref{fig:sink_ablation} ablates the sink-aware regularization weight $\beta$. Without sink suppression, \method retains a TPR of $0.946$ but incurs an FPR of $0.417$. Setting $\beta=10$ suppresses these false activations, achieving the peak mIoU of $0.774$ with a TPR of $0.927$. Further increasing $\beta$ to $12$ over-regularizes and degrades both metrics. We thus adopt $\beta=10$ as default.

\begin{table}[h]
    \centering
    \begin{tabular}{@{}lccc@{}}
        \toprule
        \textbf{Policy} & \textbf{Authorized Sources} & \textbf{TPR$\uparrow$} & \textbf{FPR$\downarrow$} \\
        \midrule
        Principal-only      & System/User instr.        & 0.949 & 0.086 \\
        Tool-authorized$^\star$ & + Tool directives       & 0.927 & 0.056 \\
        Result-authorized   & + Result sources          & 0.909 & 0.048 \\
        \bottomrule
    \end{tabular}
    \caption{Authority-policy analysis. Each row adds the corresponding source to the authority set.}
    \label{tab:policy}
\end{table}

\textbf{Configurability of the authority policy.}
Varying the authority policy (Principal-only $\rightarrow$ Tool-authorized $\rightarrow$ Result-authorized) yields stable performance, with TPR above $0.909$ and FPR below $0.086$, demonstrating that AttnLocate can adapt to configurable authority policy.

\section{conclusion}

We present AttnLocate, a runtime framework that localizes behavior-guiding instructions from attention patterns and adjudicates unauthorized invocation behaviors based on the origin of the localized spans. Extensive evaluations across ten agent configurations from six model families against tool poisoning and indirect prompt injection 
demonstrate AttnLocate's strong localization and high adjudication performance, and it can transfer to unseen models and adapt to authority policy changes without retraining. 
Our findings highlight that localizing the exact behavior-guiding instructions enables fine-grained, provenance-aware policy enforcement, a capability absent in prior detection-based or coarse-grained attribution defenses.

\bibliography{aaai2027}

% Check whether the conference requires a reproducibility checklist to be included in the paper.
% If so, you can uncomment the following line and ajust the path to include it.
% \input{ReproducibilityChecklist.tex}

\end{document}